\documentclass[sigconf]{acmart}

\usepackage{booktabs}

\setcopyright{rightsretained}

\acmConference[-]{-}{-}{-}
\acmYear{}
\copyrightyear{-}

\acmArticle{-}
\acmPrice{-}

\acmBooktitle{-}

\begin{document}
	
\title{Towards Semantically Enhanced Data Understanding}

\author{Markus Schr{\"o}der}
\affiliation{
  \institution{Smart Data \& Knowledge Services Dept., DFKI GmbH}
  \institution{Computer Science Dept., TU Kaiserslautern}
  \city{Kaiserslautern}
  \country{Germany}
}
\email{markus.schroeder@dfki.de}

\author{Christian Jilek}
\affiliation{
	\institution{Smart Data \& Knowledge Services Dept., DFKI GmbH}
	\institution{Computer Science Dept., TU Kaiserslautern}
	\city{Kaiserslautern}
	\country{Germany}
}
\email{christian.jilek@dfki.de}

\author{J{\"o}rn Hees}
\affiliation{
	\institution{Smart Data \& Knowledge Services Dept., DFKI GmbH}
	\institution{Computer Science Dept., TU Kaiserslautern}
	\city{Kaiserslautern}
	\country{Germany}
}
\email{joern.hees@dfki.de}

\author{Andreas Dengel}
\affiliation{
	\institution{Smart Data \& Knowledge Services Dept., DFKI GmbH}
	\institution{Computer Science Dept., TU Kaiserslautern}
	\city{Kaiserslautern}
	\country{Germany}
}
\email{andreas.dengel@dfki.de}

\begin{abstract}
In the field of machine learning, data understanding is the practice of getting initial insights in unknown datasets.
Such knowledge-intensive tasks require a lot of documentation, which is necessary for data scientists to grasp the meaning of the data.
Usually, documentation is separate from the data in various external documents, diagrams, spreadsheets and tools which causes considerable look up overhead. 
Moreover, other supporting applications are not able to consume and utilize such unstructured data.

That is why we propose a methodology that uses a single semantic model that interlinks data with its documentation.
Hence, data scientists are able to directly look up the connected information about the data by simply following links.
Equally, they can browse the documentation which always refers to the data.
Furthermore, the model can be used by other approaches providing additional support, like searching, comparing, integrating or visualizing data.
To showcase our approach we also demonstrate an early prototype.
\end{abstract}


\maketitle

\section{Introduction}

The purpose of data mining is to discover patterns in datasets using various methods from multiple research fields.
In order to guide data scientists in this complicated task, several process models for data mining exist \cite{kurgan2006survey}.
They all have in common that they emphasise on understanding the data before analysing it.
For instance, the popular
Cross-Industry Standard Process Model for Data Mining (CRISP-DM) \cite{chapman1999crisp} includes a ``Data Understanding'' phase in order to guide analysts in getting initial findings from data, which in particular means collecting, describing, exploring and verifying datasets.

Usually, analysts ask many questions when facing unfamiliar data in the understanding phase.
For example, they would like to know what quality issues exist, what data dependencies are known, which values occur most frequently, what columns exist, how data can be accessed, etc.
Answering those questions immediately and specifically in conjunction with the corresponding data facilitates its comprehension.
However, it requires that data scientists have to read or compose various reports describing the data.
The containing information helps them to grasp the contents of the data and to ease the incremental, repeating and time-consuming process of understanding unknown datasets.

Nevertheless, data scientists are confronted with several typical challenges in this process.
To illustrate this, we think of an imaginary data scientist who has to analyse data from some foreign company.
After the analyst begins to understand the project's objective and goals, the unknown dataset is investigated.

The heterogeneous data sources are stored using different technologies, each forcing the usage of other tools to access and browse them.
During exploration, the analyst notices that some parts of the datasets are not documented.
A couple of notations for tables, columns and values are incomprehensible because they contain technical terms, abbreviations, or identifiers.
After several iterations with his contact person in the other company, all necessary information is gathered and written down in a separate document.
But the data scientist also investigates the datasets on his own using various tools and visualizations \cite{HanKamber+2011} that produce multimedia results, like reports, lists and diagrams.

For some datasets, the company provides several documents that explain, among many other things, the meaning of the tables' columns.
But the employee needs a lot of time to find the corresponding documents and the pages explaining a certain table.
During the project, this information has to be looked up repetitively which proves to be very time-consuming, all because data and its documentation are physically and logically separated.
Additionally, because of the large amount of different kinds of documentation, the analyst loses track of which information belongs to which dataset content.
The data scientist would like to use helping tools, but the unstructured reports are rather difficult for algorithms to process.

In conclusion, the scenario shows the main problem: data and its documentation are not logically ``connected''. 
As a result, this hinders data scientists in freely moving within and between those two information spaces and consequently causes considerable overhead in the data understanding process.

\section{Approach} 

To ease the understanding of data, we would like to support data scientists with an easy way to look up corresponding documentation about unknown data.
In the following, we will step-by-step explain how our proposed methodology enables this.

A first solution would be to include the documentation into the data source.
In fact, some information systems allow to add textual comments for their data elements (e.g. PostgreSQL supports a \texttt{COMMENT} command \cite{douglas2003postgresql}).
Another approach in this regard is CCSV \cite{santos2017contextual} which combines content and context (in form of semantic data) in regular CSV files.
However, such practices have several limitations and drawbacks. 
The data source has to be modified to include comments and they only reside in the system where they are created.
As a consequence, in order to make them available for others, the whole data source has to be shared.
Often, the additional information is limited by size and restricted to a certain format (mostly textual).

Instead of locally stored annotations, we would like to save them externally.
This has the benefit that they are not limited by the data source's capabilities.
If annotations are modelled separately, they must refer to the piece of data they annotate.
Usually, analysts use labels in their reports to implicitly address data elements like tables or columns, but this causes the look up overhead which was discussed earlier.
Instead, we would like to have a simple resolvable hyperlink into tabular data.
Actually, such a deep link should be able to refer to a specific part of a table.
That way, a look up becomes as easy as browsing a web site.
This approach is comparable with URI fragment identifiers\footnote{\url{https://tools.ietf.org/html/rfc3986\#section-3.5}} which refer to subordinate resources of primary ones as known from HTML.
In the context of tabular data, fragment identifiers have been proposed for CSV files\footnote{\url{https://tools.ietf.org/html/rfc7111}}.
We already demonstrated \cite{schroeder2018deeplinker} that such deep links are a convenient way to address certain locations within desktop resources.

With this method, data scientists are now able to use hyperlinks in their reports to refer to the described data.
But, as we have already noticed, reports in free text form are rather difficult to process.
That is why our approach suggests that the documentation is (a) modelled in a machine-processible form or that (b) the documents' contents is at least similarly addressable.
The latter means that we are able to refer to parts of documents about data (like reports, presentations or spreadsheets).
With this approach it is now possible to interlink data with its documentation.
We utilize the Resource Description Framework\footnote{\url{https://www.w3.org/TR/rdf11-primer}} (RDF) which allows to link resources (identified by hyperlinks) using triples.
As a result, the look up of associated documentation reduces to a simple traversal in a semantic graph.

Although we create another separate model, this approach does not duplicate the datasets.
Often, legacy data is fully converted to RDF before it is further processed in an RDF graph.
For instance, csv2rdf\footnote{\url{https://www.w3.org/TR/csv2rdf}} suggests how tabular data should be converted.
Other approaches bypass this by generating RDF on-demand during runtime, similarly as we do.
For example, D2RQ \cite{bizer2004d2rq} focuses on making relational databases queryable by SPARQL\footnote{\url{https://www.w3.org/TR/sparql11-query}} 
without copying the database.
However, the application initially needs a mapping which defines how database contents is mapped to RDF.
In comparison, our approach only requires that the data source is addressable in order to make statements about it.
Because of such referencing, the data stays where it is and details are only requested when necessary.

Having a semantic representation of the data and its documentation yields additional modelling possibilities.
The first one is cross-referencing, which allows the reciprocal linking of two or more datasets or documents.
This is helpful when information is spread across various sources, for instance, when a CSV file and a database table stores information about the same entity which is also mentioned in a reporting document.
In addition, ontologies \cite{gruber1993translation} can be used to further model the data's domain.
The same way, we are able to link the legacy data to resources from 
DBpedia \cite{auer2007dbpedia}, which contains structured information from Wikipedia.
As an example, a column containing motorcar identifiers could link to DBpedia's car entity\footnote{\url{http://dbpedia.org/resource/Car}} to clarify its content.
Such approaches are used in Semantic Data Mining \cite{Dou2015}: a research field that incorporates domain knowledge into usual data mining processes.

The modelled RDF data is queryable with SPARQL 
which already covers many use cases. 
To give an example: now data scientists are able to list columns from several CSV files that have only one distinct value. 
Other supporting approaches can make use of the semantic graph, too.
Search engines that index all datasets and documentation retrieve semantic entities based on a keyword-based search.
In this context, thanks to the modelling, semantic search \cite{guha2003semantic} also becomes possible.
Besides, data visualization can make use of the additional meta data.
Based on the data properties, approaches can automatically suggest more appropriate depictions using the right diagram type.
In this regard, it always remains possible to generate artefacts like reports or plots from the semantic data.
Especially reporting documents are well-known and understood exchange mediums in the data science community.
\\
\\
To sum up, our contribution is the combination of mentioned solutions to formulate a methodology that uses a semantic model which interlinks addressable data with its documentation.
That way, data scientists interact with one model allowing for an easy look up of required information to understand data.
Such models are commonly classified as data catalogues\footnote{Also known as or comparable with database catalogue, data dictionary/directory or meta data repository.} which store meta data about dataset contents like tables and columns.
We describe our specialized model as a ``semantic catalogue of data made addressable'' to emphasise that our catalogue semantically enriches referred data.

\begin{figure}[h]
	\includegraphics[width=\linewidth]{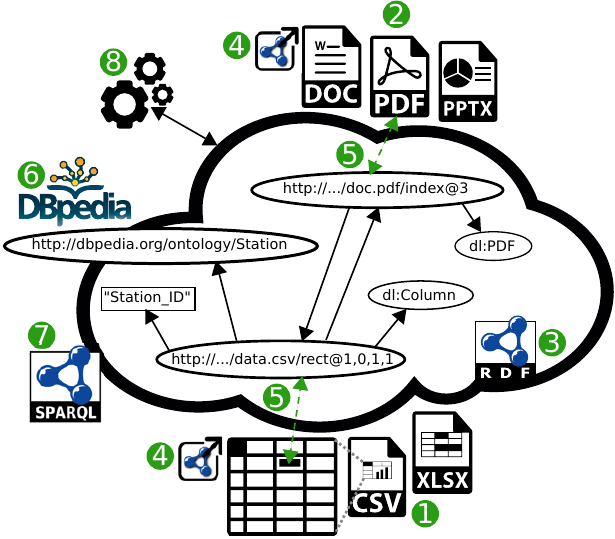}
	\caption{\textcircled{\small{1}} Data, \textcircled{\small{2}} Documentation, \textcircled{\small{3}} RDF Graph, \textcircled{\small{4}} Tool for \textcircled{\small{5}} Deep Linking, \textcircled{\small{6}} External Knowledge Base, \textcircled{\small{7}} Querying, \textcircled{\small{8}} Supporting Approaches.}
	\label{fig:overview}
\end{figure}

\pagebreak
\autoref{fig:overview} exemplary illustrates our approach and the relationships of the various solutions.
Data \textcircled{\small{1}} and documentation \textcircled{\small{2}} are interlinked using an RDF graph \textcircled{\small{3}}.
With our specialized tool \textcircled{\small{4}} data scientists can browse and annotate table contents as well as documents.
Deep links \textcircled{\small{5}} are used to identify and refer to specific data and documentation parts.
The semantic graph interconnects and further enriches them with class types, literals and resources from external knowledge bases \textcircled{\small{6}}.
SPARQL \textcircled{\small{7}} enables querying the graph while other approaches \textcircled{\small{8}} are now able to consume and contribute to it.

There are also few remaining challenges.
Our approach provides that a specialized tool (see next section) is used to semantically browse and annotate datasets.
To aid data scientists in becoming familiar with the proposed methodology, we will make use of our previously developed approaches \cite{schroeder2018spreadsheet, schroeder2018sparql}.

\section{Implementation}

To showcase our approach we demonstrate an early prototype which is an extension of DeepLinker \cite{schroeder2018deeplinker}. 
For better demonstration, an online version is available\footnote{\url{http://www.dfki.uni-kl.de/~mschroeder/demo/deeplinker-data-understanding}}.

Our locally running server application allows to browse a table-based file (CSV or Excel) in a web browser. 
The application responds with an HTML representation of the resource along with links pointing deeper into the table.
As a result, instead of just stopping at the resource's content (surface link), users can traverse further into them.
The resulting deep links allow users to point to any cell or region which enables to make statements about them in form of RDF triples.

\begin{figure}[h]
	\includegraphics[width=\linewidth]{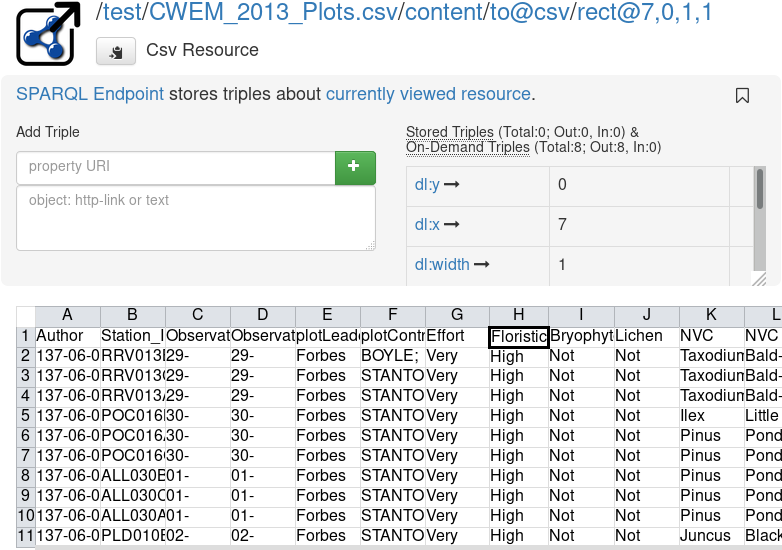}
	\caption{A page which shows a selected cell (H1) of a CSV resource together with a form to add and list RDF triples about this cell (highlighted in gray).}
	\label{fig:deeplinking}
\end{figure}

For demonstration purposes, we randomly chose a dataset about vegetation plots\footnote{\url{https://ecos.fws.gov/ServCat/Reference/Profile/44947}} provided by the Environmental Conservation Online System (ECOS) via the U.S. Government open data platform\footnote{\url{https://www.data.gov}}.
\autoref{fig:deeplinking} shows a screen shot of our tool which browses the file \texttt{CWEM\_2013\_Plots.csv}\footnote{\url{https://ecos.fws.gov/ServCat/DownloadFile/125253}} while cell H1 is selected.
In general, the page shows the accessed link (top) and a simple form to add and list RDF triples (highlighted in gray).
Below, an excerpt of the table together with the selected cell is rendered.
Note that the column and row headers as well as each cell is again a deep link.

Since we would like to demonstrate how analysts would interact with documented data, we implemented some automated analyses that scan the data and generate RDF statements.
Having columns and their values located in the table, it counts the total, distinct, blank, and empty values. 
Descriptive statistics, such as minimum, average, standard deviation and maximum, are created for the values' lengths.
A histogram for distinct values is generated showing their frequency distribution.

Every insight mentioned above is formulated as one or more RDF statements.
The resolvable deep links are used at the triples' \textit{subject} and \textit{object} positions and identify and refer to the table's contents.
We plan to utilize other existing approaches that generate analogous results of analysis, like \cite{Vesanto} or \cite{Sukhobok2017}, by adjusting them to generate similar RDF data.
Besides automatically generated statements, data scientists may add further annotations while exploring the data.
For that, DeepLinker offers a form to add RDF statements about the currently browsed resource by providing a simple vocabulary (classes and properties).
An even easier RDF data entry method could be provided by using the spreadsheet metaphor \cite{schroeder2018spreadsheet}.

\begin{figure}[h]
	\includegraphics[width=\linewidth]{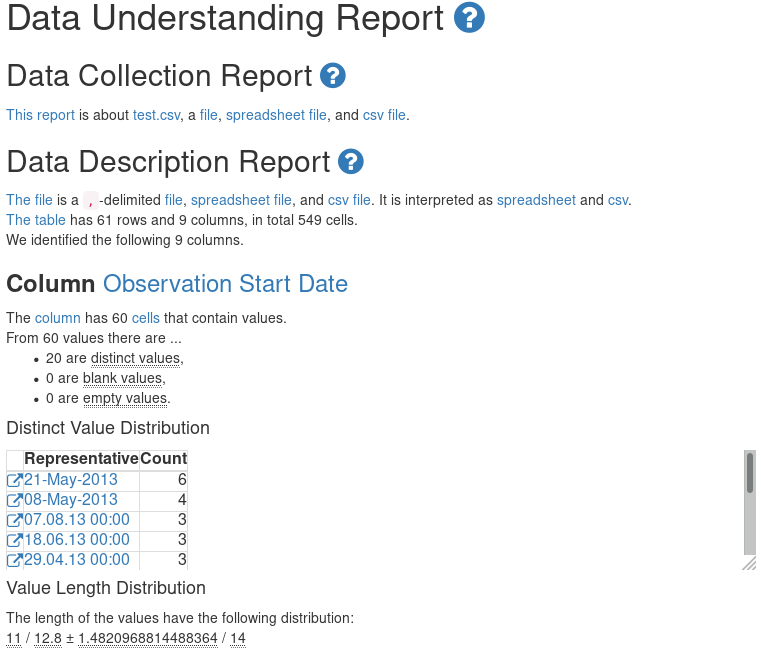}
	\caption{The report's HTML page which is generated based on the analysed data.}
	\label{fig:reporting}
\end{figure}

\pagebreak
Since an RDF graph is rather difficult for data scientists to read and understand, our tool renders a familiar reporting document based on it (\autoref{fig:reporting}).
A template language\footnote{\url{https://freemarker.apache.org/}} is used that queries the semantic graph and outputs HTML.
The pages heavily uses deep link anchors in order to directly refer to the data.
Describing tool tips are shown when mouse hovers.

\section{Conclusion and Outlook}

In this paper, we identified several challenges in the process of understanding data.
Our proposed solution suggests the usage of a semantic model that interlinks data with its documentation, both of which were made addressable before via our deep linking approach.
We argued that this enables a simple look up of expressive information about unknown data which facilitates its comprehension.
Several additional benefits of our approach were discussed. 
Some were also demonstrated in an online\footnote{\url{http://www.dfki.uni-kl.de/~mschroeder/demo/deeplinker-data-understanding}} prototype.

In the future, we plan to conduct a user study to measure the impact of our methodology in real world scenarios.
Looking at other data science tasks, we would like to use the acquired semantic graph to also enhance business understanding, data preparation, modelling and evaluation.
We are confident that our approach enables new possibilities to assist in Semantic Data Mining, a research field that incorporates domain knowledge into usual data mining processes.

\vspace{3cm}
\bibliographystyle{acmart}
\bibliography{paper}

\end{document}